\documentclass[referee]{raa}
\usepackage{graphicx,times}
\usepackage{natbib}
\usepackage{amssymb,amsmath}
\usepackage{subfigure}
\bibpunct{(}{)}{;}{a}{}{,}

\usepackage[a4paper=true,dvipdfm=true,pagebackref=true]{hyperref}
\hypersetup{pdftitle = The title of my PDF, pdfauthor = My name, pdfsubject= The subject, pdfkeywords = keyword1 keyword2 keyword3}
\hypersetup{colorlinks = true, linkcolor = green, anchorcolor = red, citecolor = blue, filecolor = red, pagecolor = red, urlcolor = red}

\begin{document}

   \title{A high-contrast coronagraph for earth-like exoplanet direct imaging: design and test
$^*$
\footnotetext{\small $*$ Supported by the National Natural Science Foundation of China and Strategic Priority Research Program, CAS.}
}

 \volnopage{ {\bf 2012} Vol.\ {\bf X} No. {\bf XX}, 000--000}
   \setcounter{page}{1}

   \author{Cheng-Chao Liu \inst{1,2,4}, De-Qing Ren \inst{1,2,3}, Jian-Pei Dou\inst{1,2}, Yong-Tian Zhu
      \inst{1,2},  Xi Zhang \inst{1,2}, Gang Zhao \inst{1,2}, Zhen Wu \inst{1,2}, Rui Chen \inst{1,2}
   }

   \institute{ National Astronomical Observatories$/$Nanjing Institute of Astronomical Optics $\&$ Technology,
Chinese Academy of Sciences, Nanjing 210042,China; {\it ccliu@niaot.ac.cn}\\
        \and
             Key Laboratory of Astronomical Optics $\&$ Technology, National Astronomical Observatories$/$Nanjing Institute of Astronomical Optics $\&$ Technology, Chinese Academy of Sciences, Nanjing 210042,China\\
	\and
Physics $\&$ Astronomy Department, California State University Northridge, 18111 Nordhoff Street, Northridge, California 91330-8268, USA\\
\and
University of Chinese Academy of Sciences, Beijing 100049, China\\
\vs \no
}

\abstract{The high-contrast coronagraph for direct imaging earth-like exoplanet at the visible needs a contrast of $10^{-10}$ at a small angular separation of $4\lambda/D$ or less. Here we report our recent laboratory experiment that is close to the limits. The test of the high-contrast imaging coronagraph is based on our step-transmission apodized filter. To achieve the goal, we use a liquid crystal array (LCA) as a phase corrector to create a dark hole based on our dedicated focal dark algorithm. We have suppressed the diffracted and speckle noise near the star point image to a level of 1.68$\times$$10^{-9}$ at $4\lambda/D$, which can be immediately used for the direct imaging of Jupiter like exoplanets. This demonstrates that high-contrast coronagraph telescope in space has the potentiality to detect and characterize earth-like planets.
}

   \authorrunning{C.-C. Liu et al. }            
   \titlerunning{A high-contrast coronagraph for Earth-like exoplanet direct imaging}  
   \maketitle

%
\section{Introduction}           
\label{sect:intro}

Today, searching the universe for earth-like exoplanets is one of the hottest topics in science. Up to now more than 1,000 exoplanets have been detected and confirmed. Most of them were discovered through indirect ways such as the transit method and radial velocity detection. These methods have not been proven sensitive enough for the detection of earth-like planets. The Kepler mission has discovered 2 earth-size planets: Kelper-62 in the habitable zone and Kepler-78b by transit approach recently (\citealt{Borucki+etal+2013};\citealt{Pepe+etal+2013}). However, direct imaging is needed to fully detect the signatures of life or biomarkers $O_2$ and $O_3$ at visible wavelengths on the exoplanets by using spectroscopic techniques. To realize this goal, the unwanted diffraction light from host star and speckle noise from the optics system must be efficiently suppressed using high-contrast imaging coronagraphs. A contrast of $\sim10^{-10}$ is required for imaging earth-like planets in the visible wavelength, which is a great challenge for current techniques. A critical issue is the speckle noise which is induced by any optical imperfection, such as the wave-front error and surface roughness. The static speckle noise, in generally degrades the contrast of a coronagraph to a level on the order of $10^{-5}\sim10^{-7}$ (\citealt{Dou+etal+2010}; \citealt{Ren+etal+2010}).

A common approach used to achieve this high-contrast by other research groups is to correct the optical errors and generate a small local dark hole by using one or two deformable mirrors (DMs), such as the groups led by Jet Propulsion Laboratory (JPL) and Princeton (\citealt{Trauger+Traub+2007}; \citealt{Kay+etal+2009}). However, this approach needs a deformable mirror (DM) with a large number of actuators and complicated phase retrieval algorithm. Although the contrast can reach the level of $10^{-10}$, the region of the dark holes created by this two groups is not large enough to apply to space telescopes for exoplanet detecting. In our group, we have successfully used the stochastic parallel gradient descent (SPGD) algorithm to efficiently supress speckle noises according to the point spread function (PSF) on the imaging focal plane, and this algorithm is first used in our extreme adaptive optics (ExAO) for a ground-based high-contrast system, which successfully demonstrated that a small dark hole with an extra contrast gain of 65-time improvement is achievable with a DM of 140 actuators (\citealt{Ren+etal+2012}).  Recently, this approach is used in our coronagraph, and the dark hole created by our approach is much larger than that created by other groups, which we will discuss in Section 3.

Ren and Zhu have developed a stepped transmission filter-based coronagraph (\citealt{Ren+Zhu+2007}). Recently new designed filter can realize a contrast of $10^{-6}$ at a small angular separation down to $4\lambda/D$ ($\lambda$ is the wavelength and D is the diameter of telescope aperture) in our laboratory test.

Based on the stepped filters, test results of the new designed coronagraph, combined both the stepped transmission filters and  our SPGD based dark-hole correction, are presented in Section 2. In Section 3 we show the test optics system and experimental results. Finally, we present the conclusions and outline future developments.
\section{TESTING RESULTS OF NEW STEPPED FILTERS AND THE DARK-HOLE CORRECTION BASED ON SPGD ALGORITHM}
\label{sect:Obs}
\subsection{New filters and the testing results}
Ren $\&$ Zhu (2007) have firstly proposed a coronagraph based on stepped transmission filters. This kind of coronagraph can be applied to an off-axis space-based telescope which is aimed for a future space mission for direct imaging of earth-like exoplanet. In our previous papers, we have demonstrated this feasibility with a high-contrast imaging coronagraph (\citealt{Dou+etal+2009}). Recently we have designed and manufactured five new filters with the goal for better contrast over a large discovery area.

Here we briefly give the design and testing results in our laboratory. For each filter, the diffraction is suppressed along one direction only (x or y), depending on the orientation of the transmission attenuation in the filter. On the coronagraph focal plane, the complex amplitude of the electric field is the Fourier transform of the complex amplitude of the pupil. The theoretical designed contrast for one filter should reach $10^{-6}$ at the angular distance of equal or larger than $4\lambda/D$ from the PSF intensity peak. For this design, two filters can also be combined together one after the other with $90^\circ$ orientation. In this case diffraction along the two perpendicular directions of x and y are suppressed, and a theoretical contrast of $10^{-8}$ should be achieved at an angular distance of $4\lambda/D$ or larger.

However, in practice the coronagraph can't achieve the theoretical contrast. Because the transmission of the filter is realized by uneven metallic coating, which introduces a thickness variation of optical path, it results in an unexpected phase. High contrast of this transmission filter based coronagraph is achieved by the so-called apodized pupil. Due to the transmission apodized pupil, the transmission throughput for each filter is 0.498. Figure 1 shows the theoretical transmission patterns of the designed filters (top panel) and the associated PSF images (bottom panels). The top panel shows images of one filter (in the left-top panel), another filter with $90^\circ$ orientation (in the central-top panel), and the combination of these two filters (in the right-top panel). In figure 2, the theoretical stepped transmission of one transmission filter and real photograph of the two orthogonal-orientation filters are presented. PSF images of coronagraph based on the combination of these two orthogonal-orientation filters under different exposure time of 0.15s, 15s and 1500s and associated contrast plot are shown in Figure 3. A 16-bit SBIG camera is used to take these images.
\subsection{The dark-hole correction based on SPGD using LCA}
\begin{figure}
   \centering
   \includegraphics[scale=0.24]{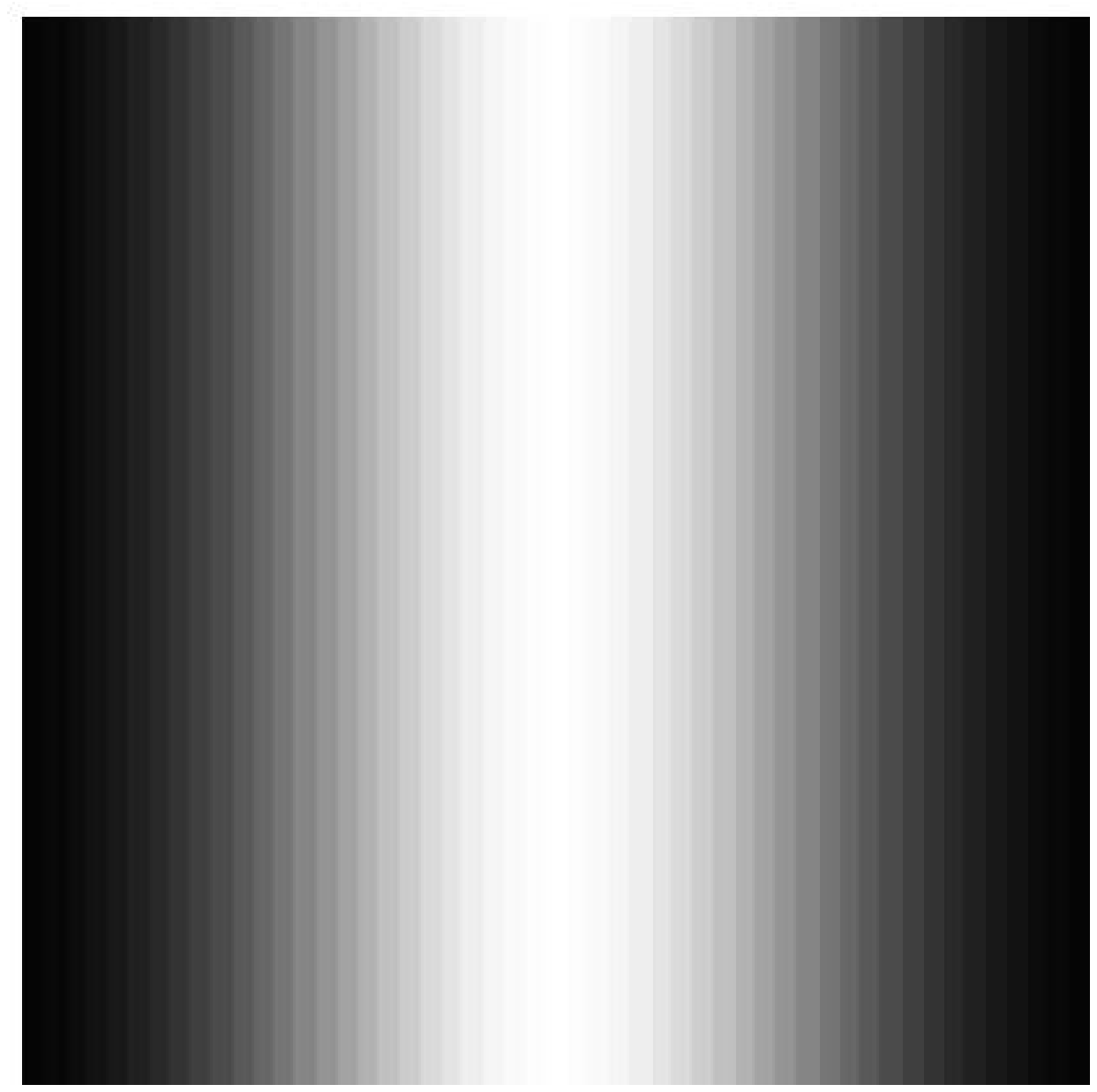}
   \includegraphics[scale=0.24]{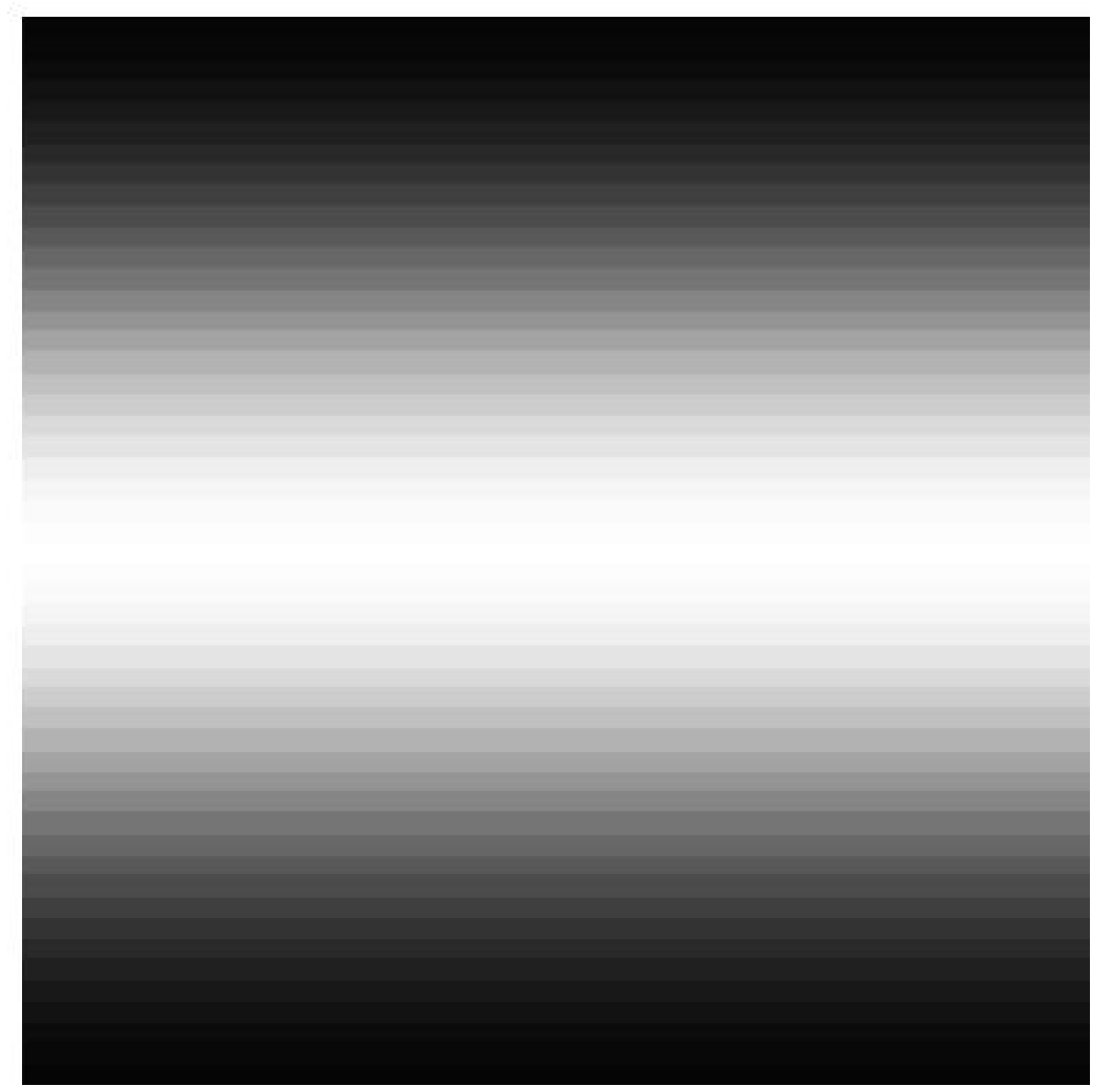}
   \includegraphics[scale=0.24]{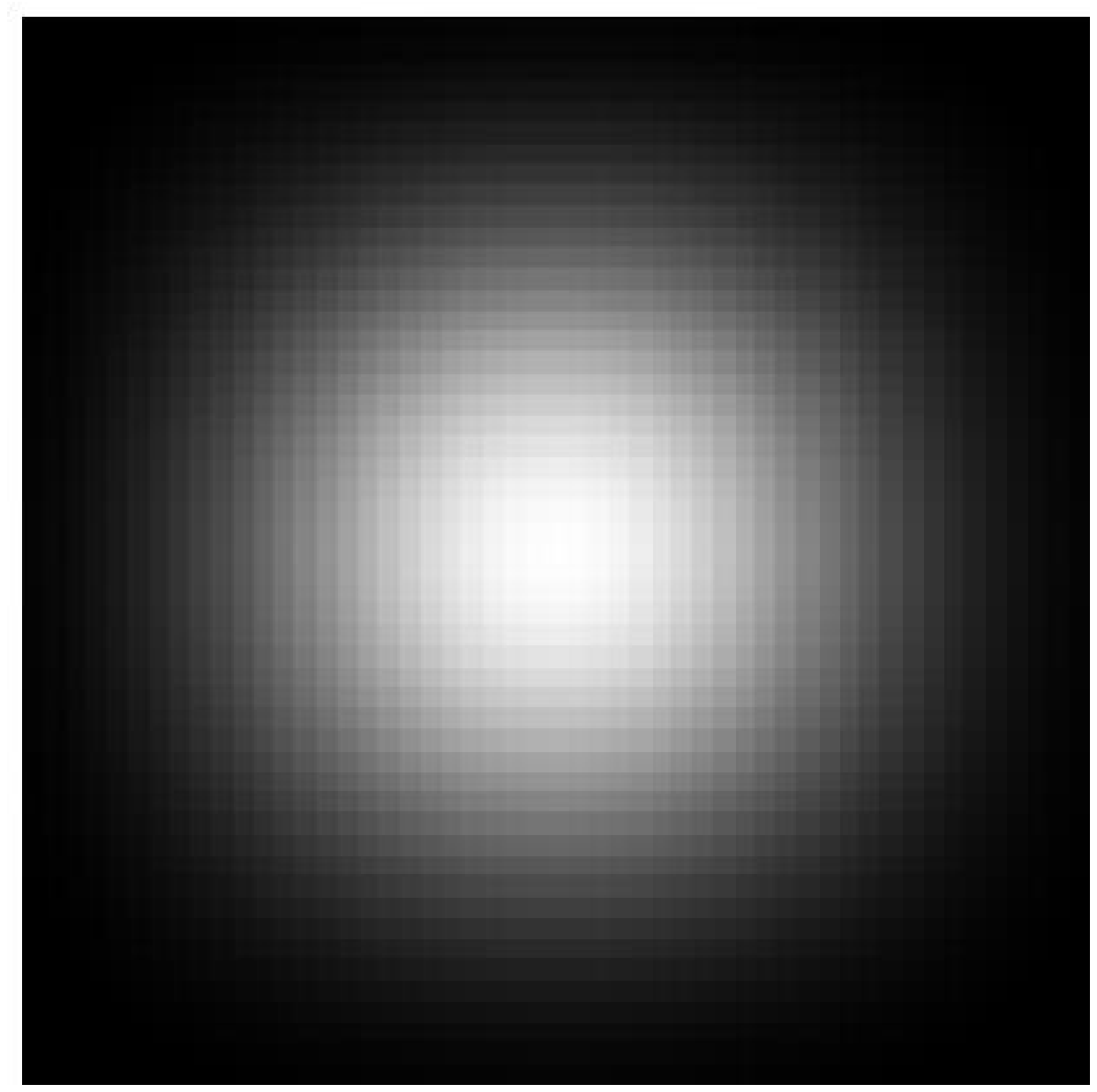}
   \centering
   \includegraphics[scale=0.24]{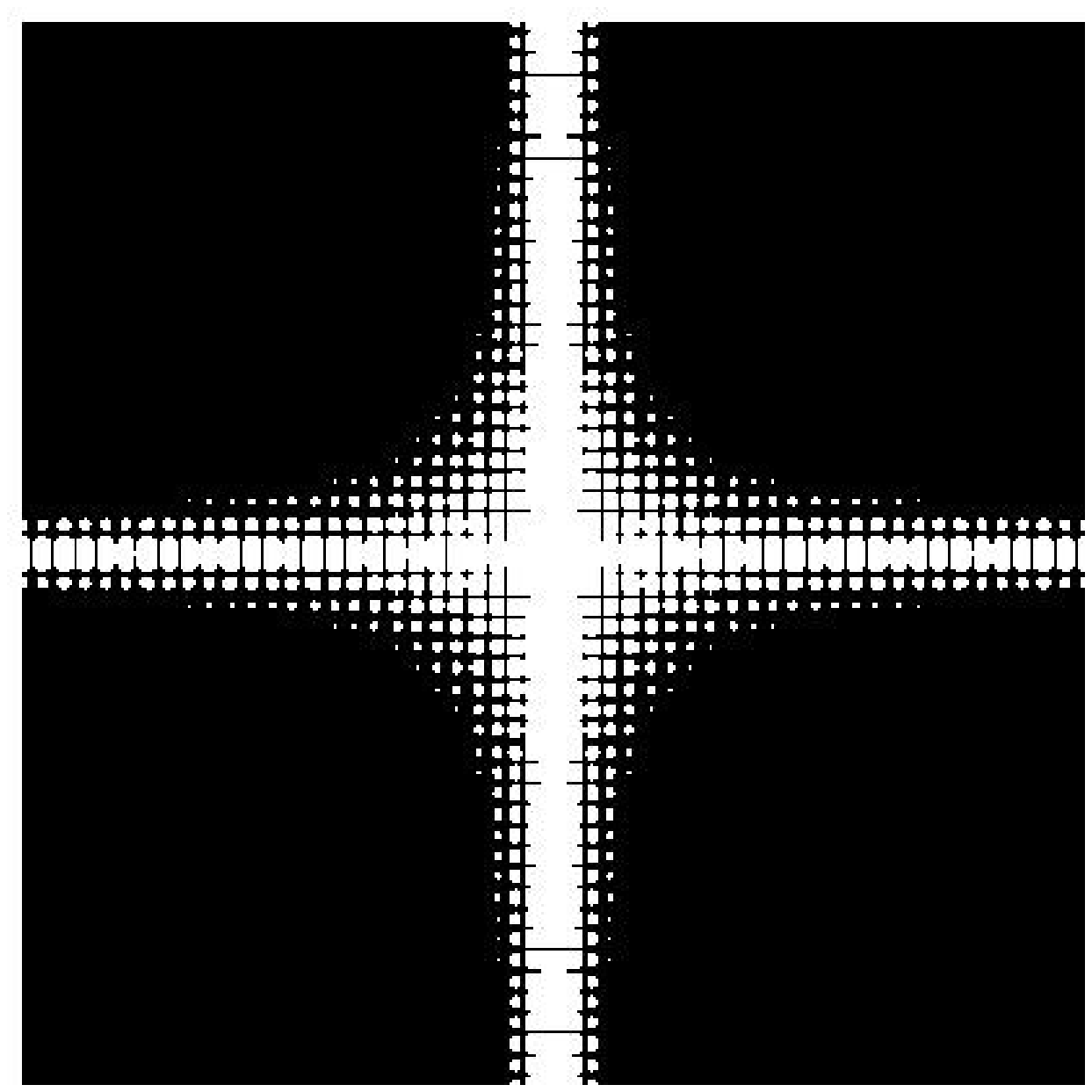}
   \includegraphics[scale=0.24]{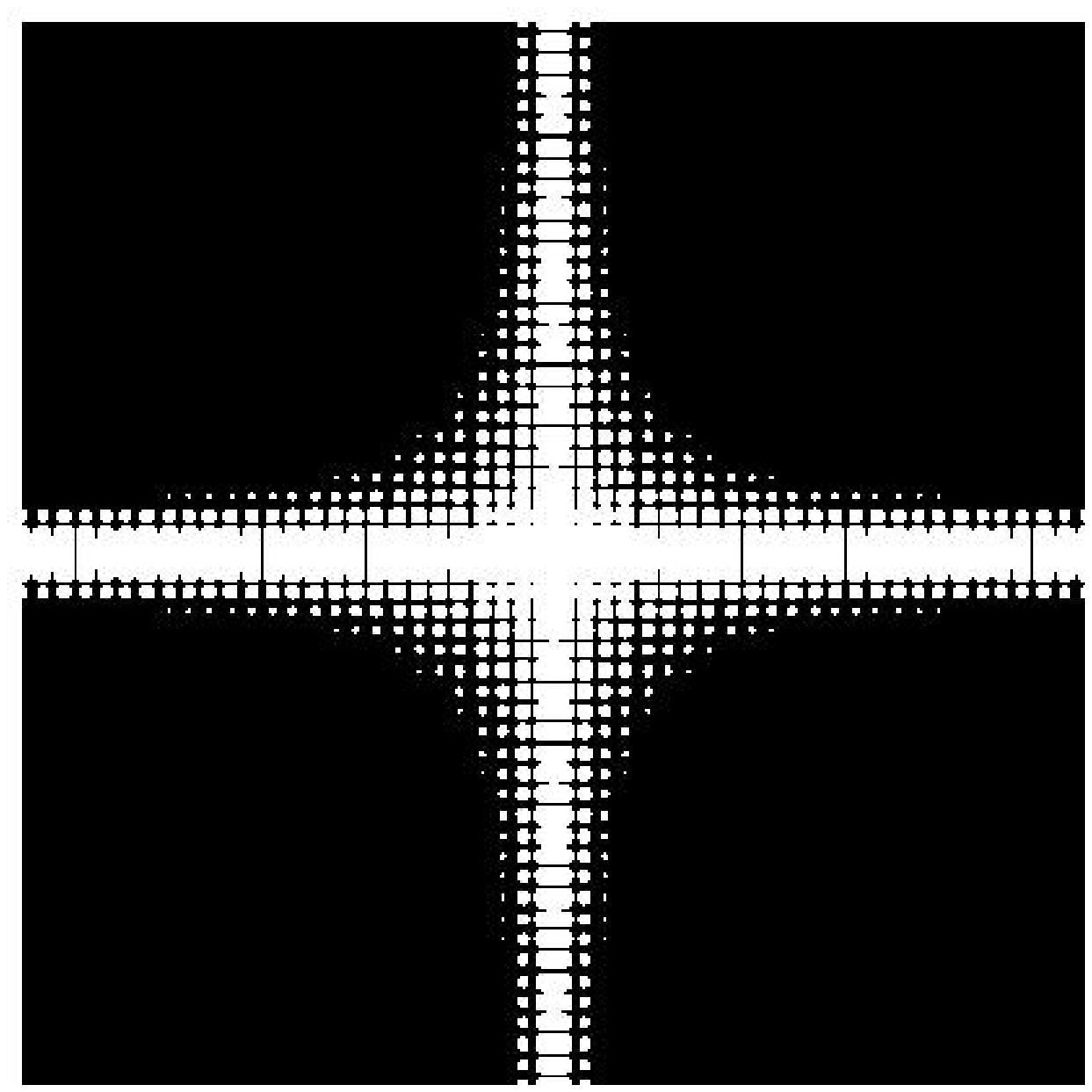}
   \includegraphics[scale=0.24]{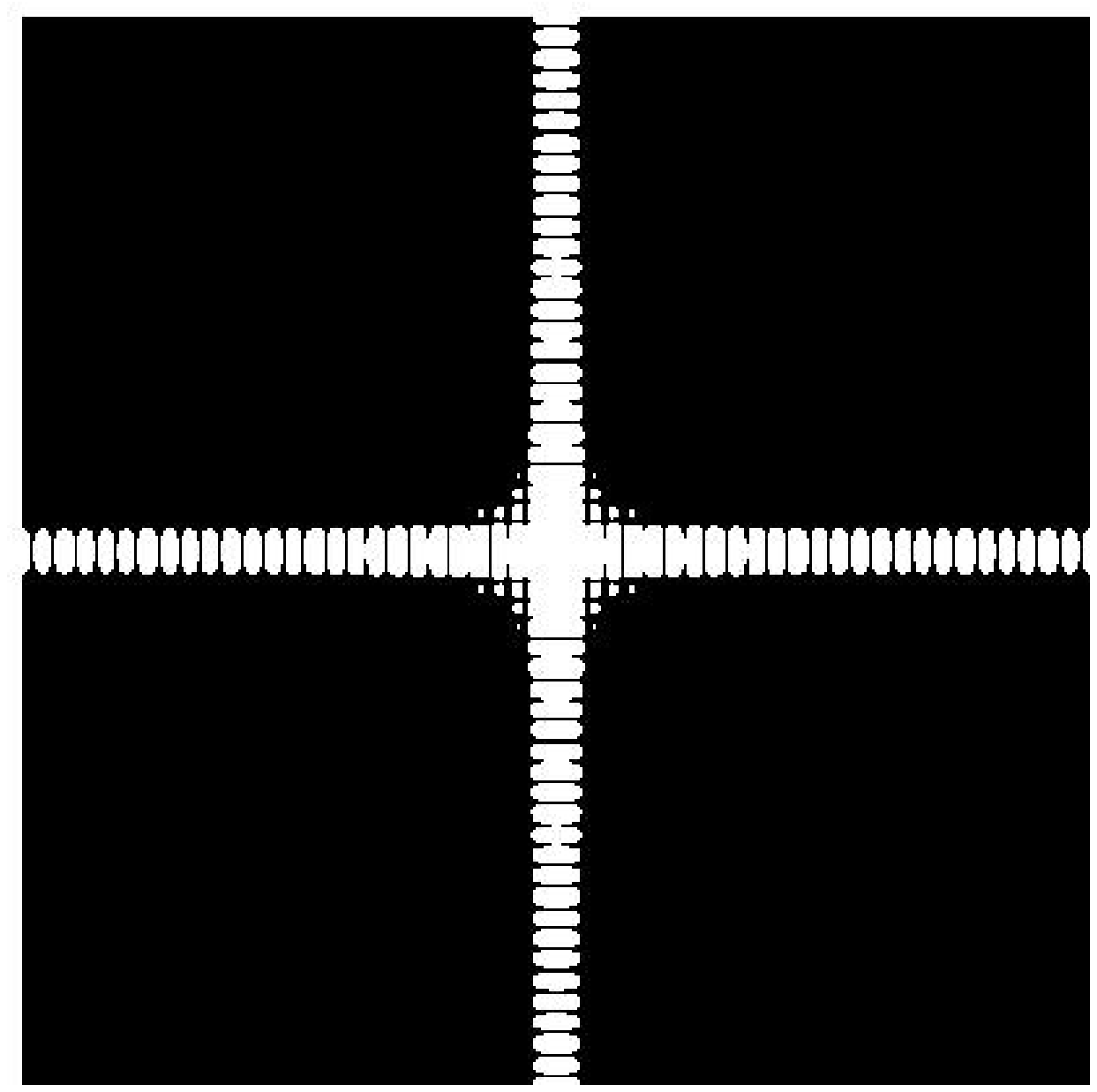}
   \caption{Transmission patterns of the filters (top) and the associated PSFs (bottom). }
   \label{Fig1}
   \end{figure}
We have firstly proposed and demonstrated that SPGD algorithm can be applied to the high-contrast coronagraph to reduce speckle noise induced by imperfection optics (\citealt{Dong+etal+2011}). In previous paper, we used a deformable mirror (DM) to provide an optimal phase to create a dark-hole in a target region and the principle of SPGD algorithm has been discussed in details somewhere (\citealt{Ren+etal+2012}). The DM used in his previous test has 140 effective actuators arranged in 12$\times$12 grid. The dark-hole has a small area because of the small actuator numbers. Considering a DM (32$\times$32 or 64$\times$64 actuators) is too expensive to afford for our tests, in our recent test, we use XY nematic series spatial light modulator (SLM) as a phase corrector to replace a DM. The SLM is  manufactured by BNS has 512$\times$512 pixels, with a 83.4\% fill factor. In the test the pixels are divided into a number of groups and each group is binned as an effective pixel consisting of 8$\times$8 pixels. The XY nematic series SLM is optimized to provide a full wave ($2\pi$) of phase stroke upon reflection at the nominal design wavelength.

In this test, we use SPGD optimization algorithm, which measures the wave-front error according to the PSF on the focal plane, and the measured information is used to directly control the phase of the LCA to create a dark-hole.
\begin{figure}
   \centering
  \includegraphics[scale=0.36]{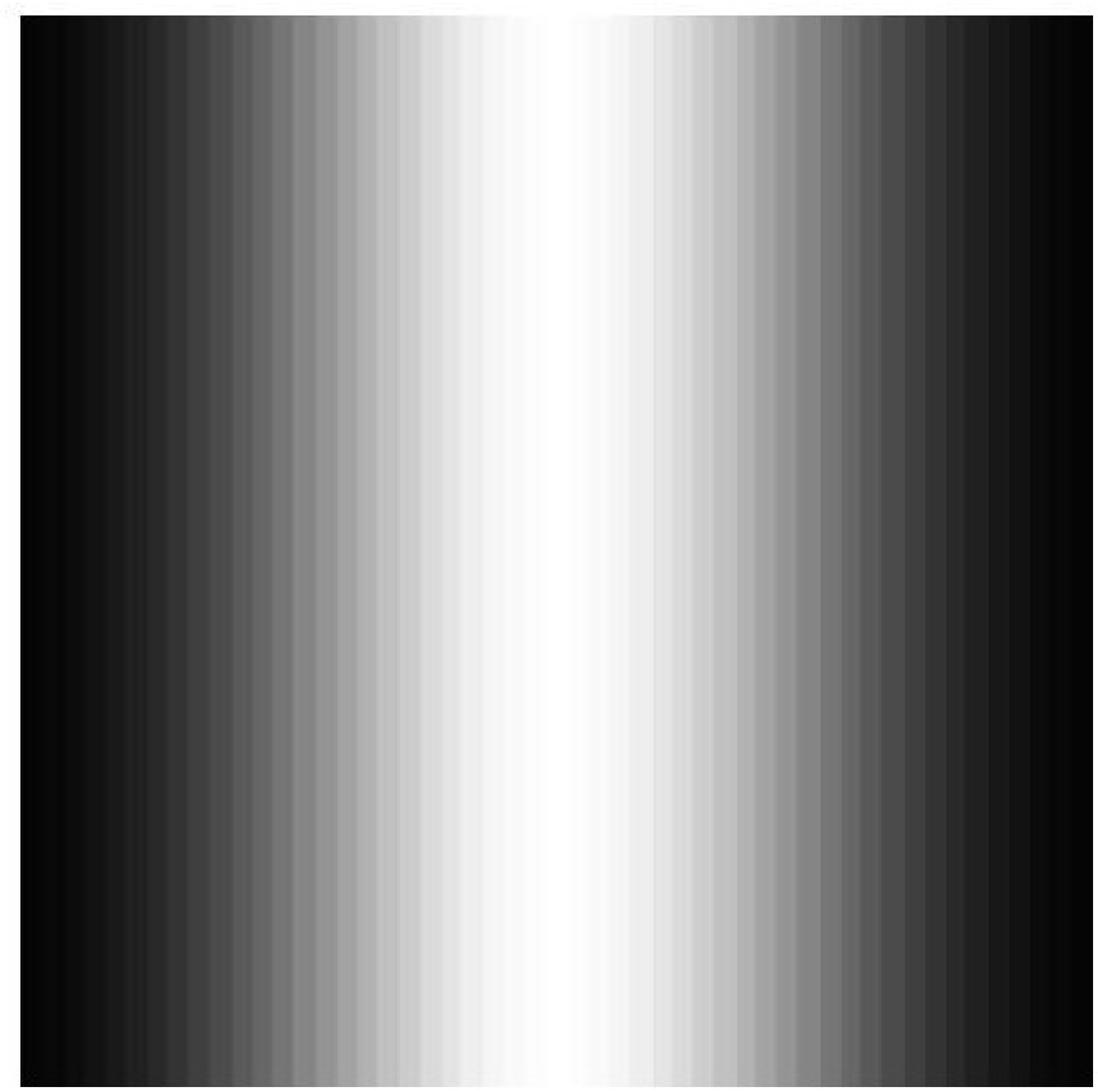}
  \includegraphics[scale=0.36]{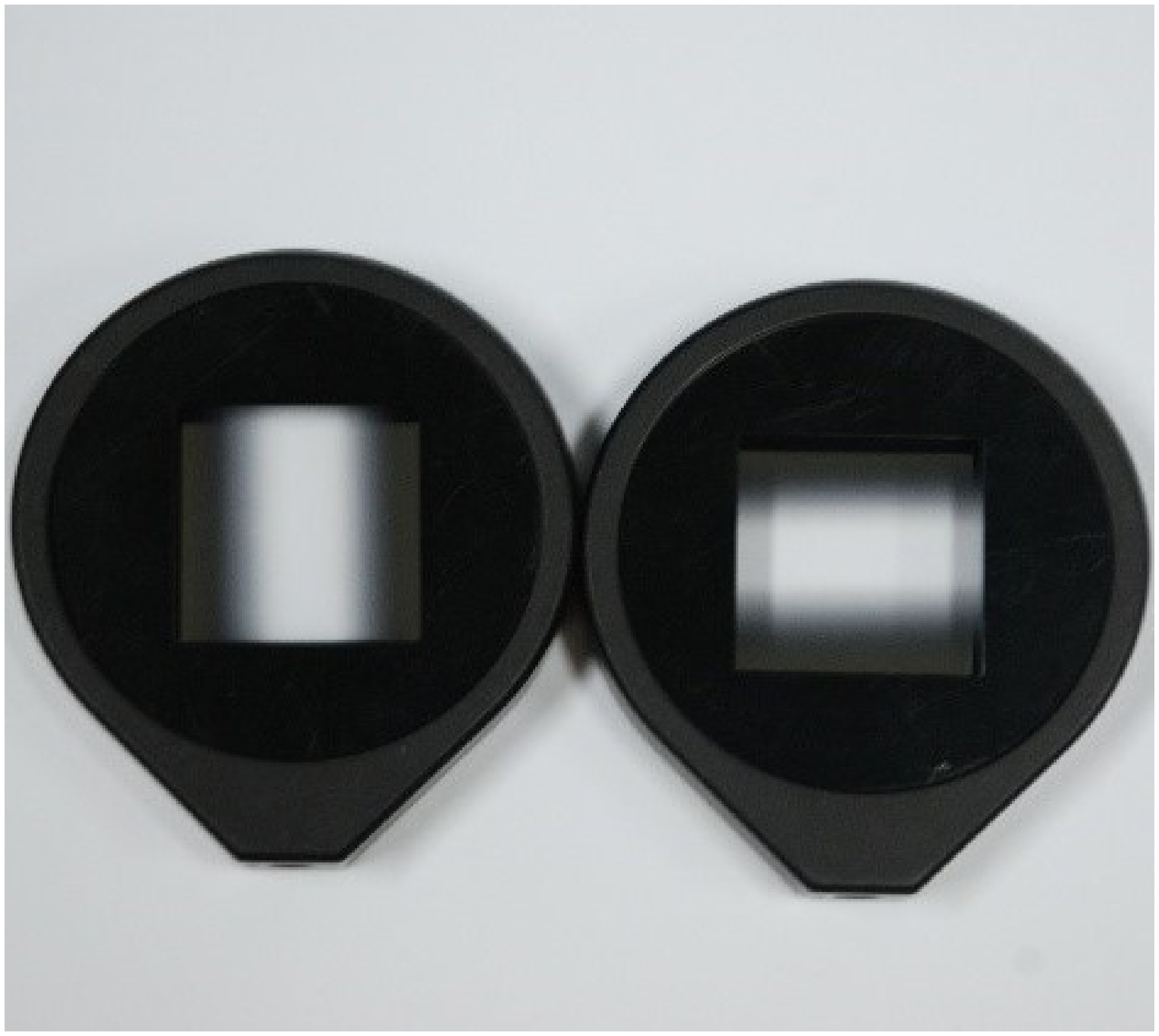}
   \caption{The theoretical 51 step transmission filter and photograph of the real two filters.}
   \label{Fig2}
   \end{figure}
   \begin{figure}
   \centering
  \includegraphics[scale=0.24]{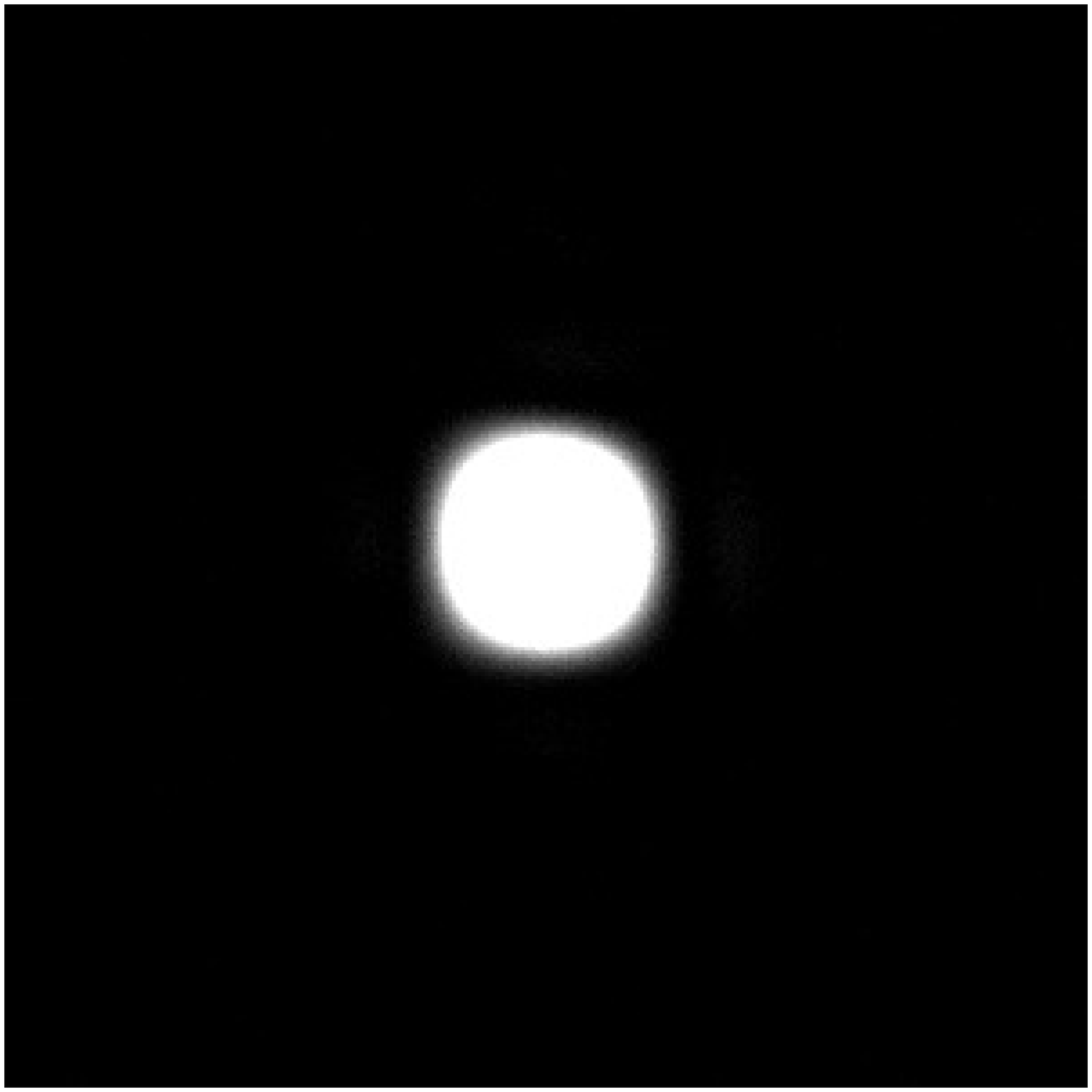}
  \includegraphics[scale=0.24]{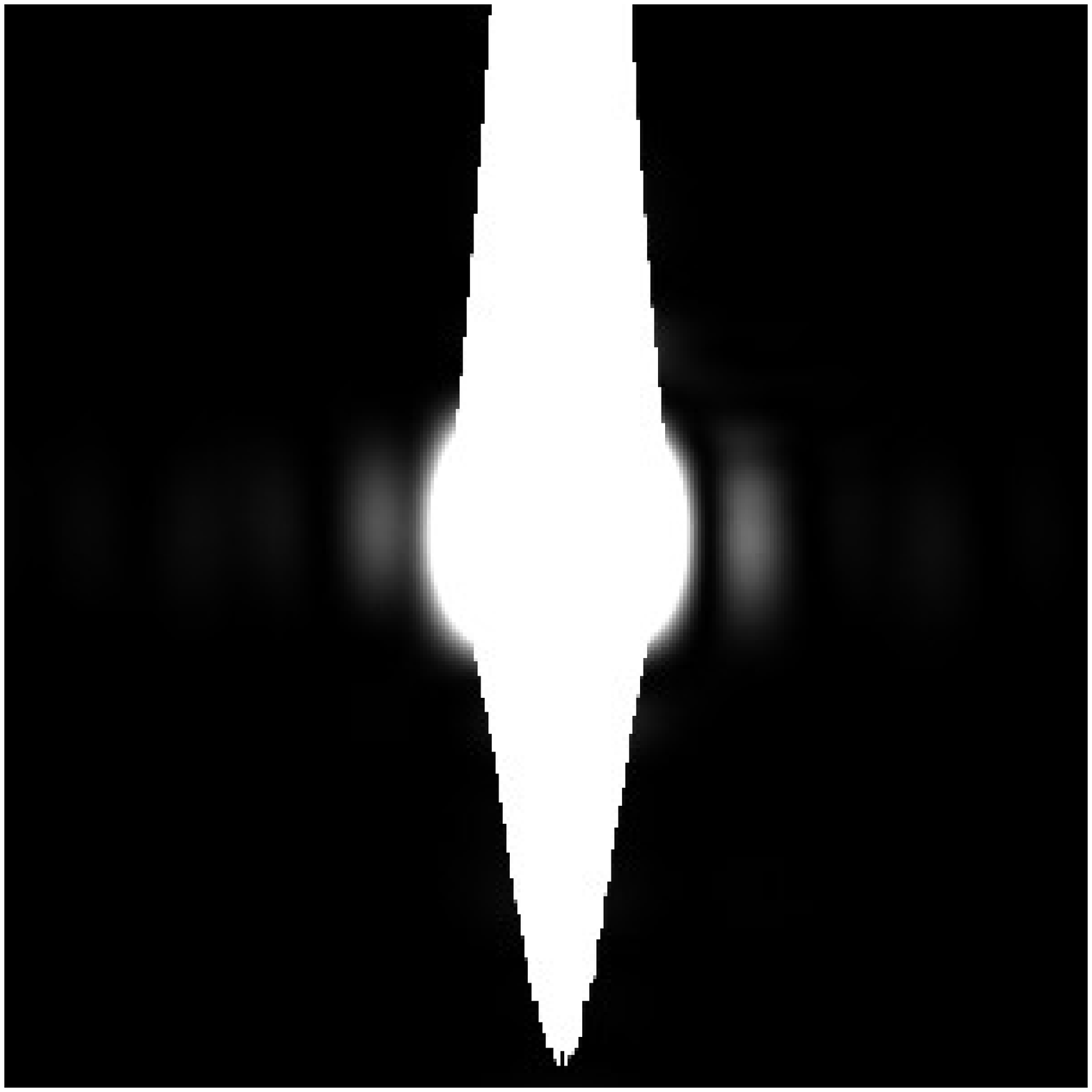}
  \includegraphics[scale=0.24]{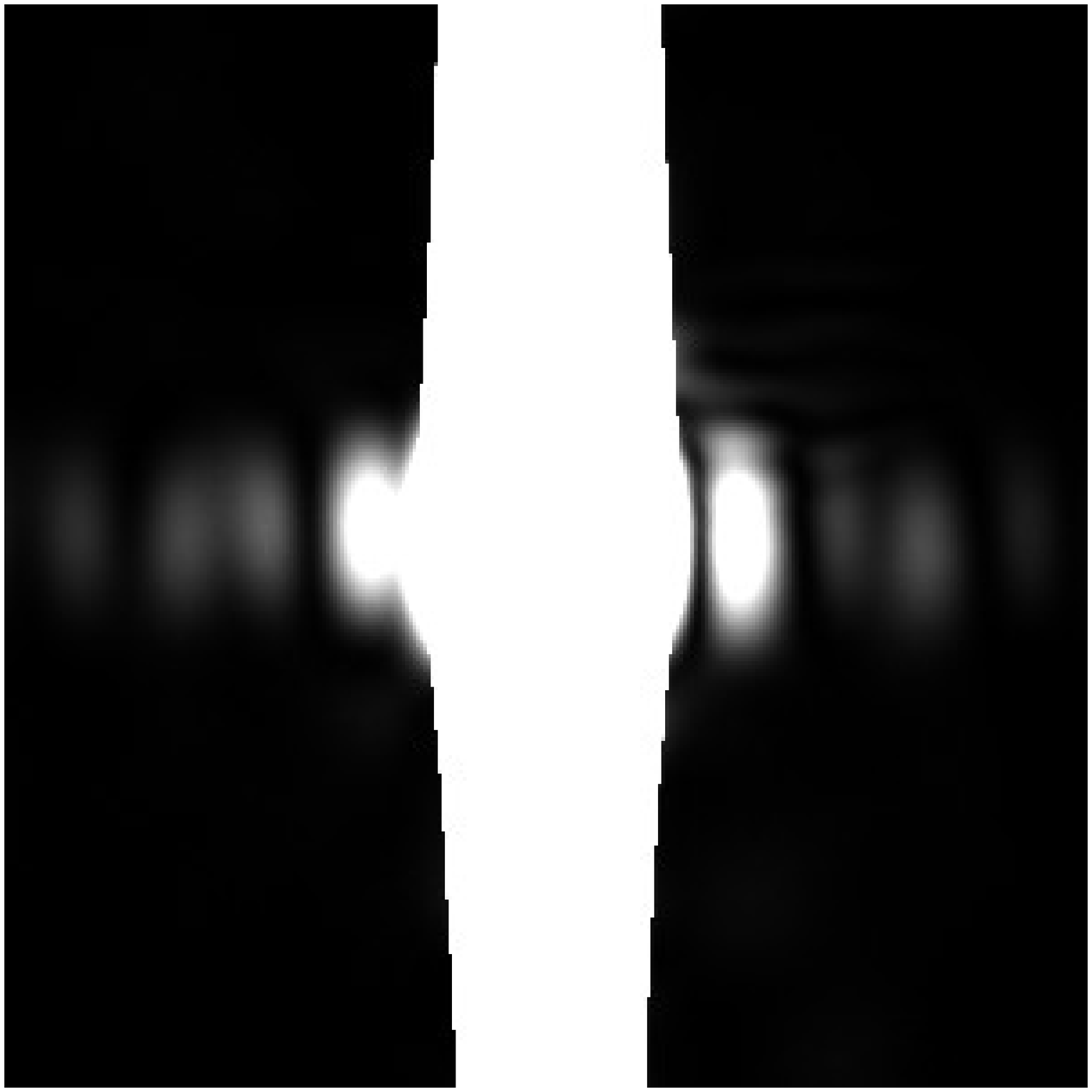}
  \centering
  \includegraphics[scale=0.72]{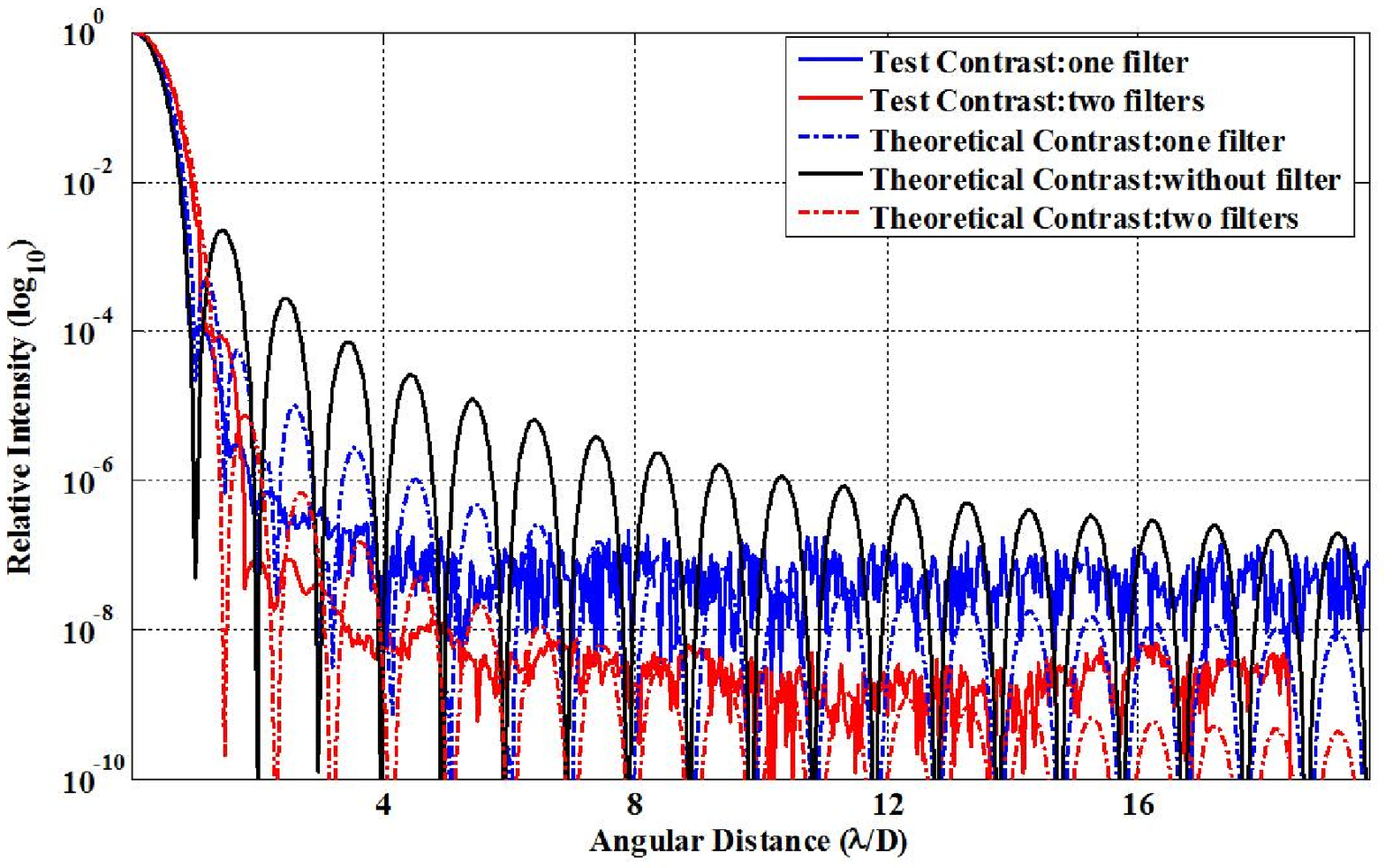}
   \caption{The top panels show the PSF images of the two filters coronagraph under different exposure time. The bright vertical pattern in central and right panels is due to the CCD image bloom. The bottom panel shows the comparison between the test and designed PSF contrast along the diagonal direction.}
   \label{Fig3}
   \end{figure}
   \begin{figure}
   \centering
  \includegraphics[scale=0.72]{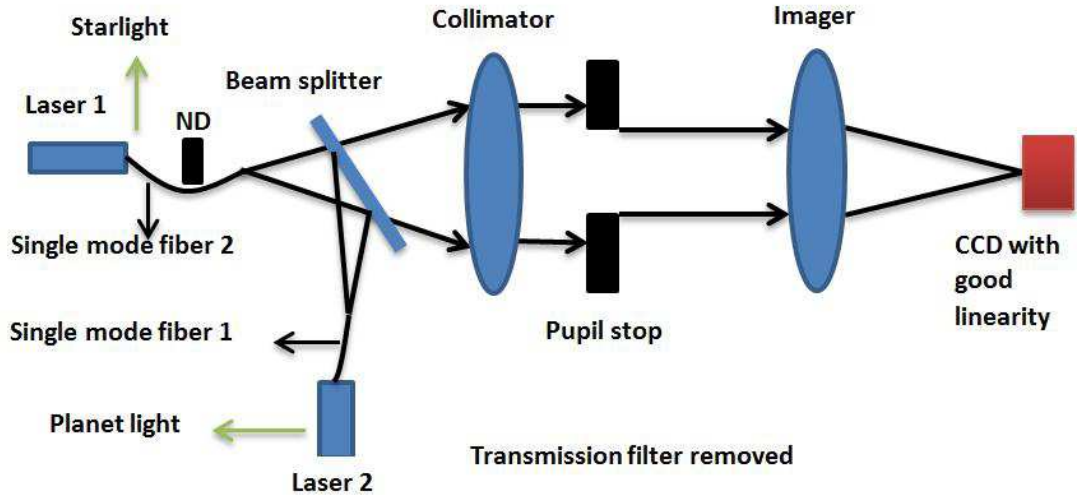}
   \caption{The optical configuration of intensity calibration.}
   \label{Fig4}
   \end{figure}
      \begin{figure}
   \centering
  \includegraphics[scale=0.36]{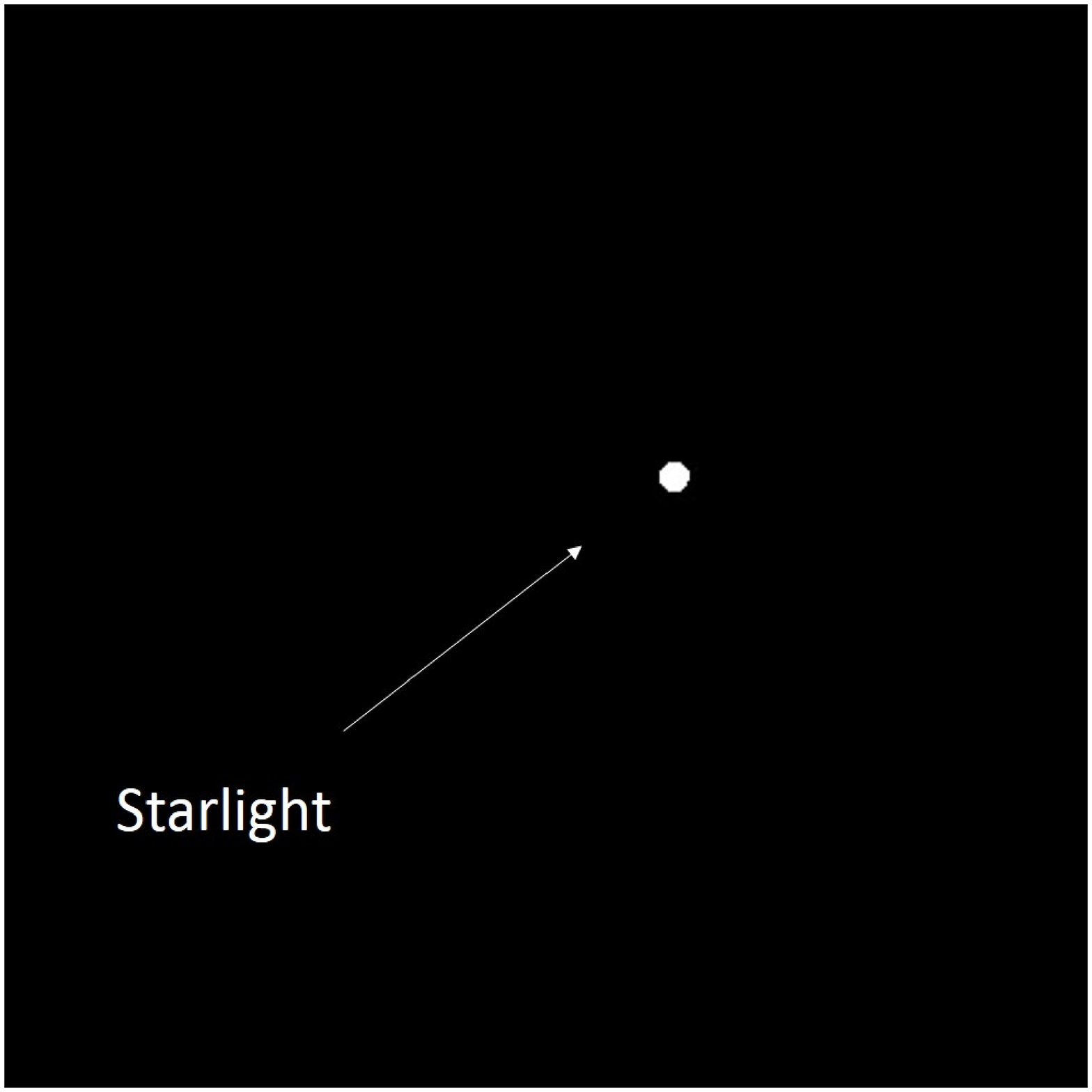}
  \includegraphics[scale=0.36]{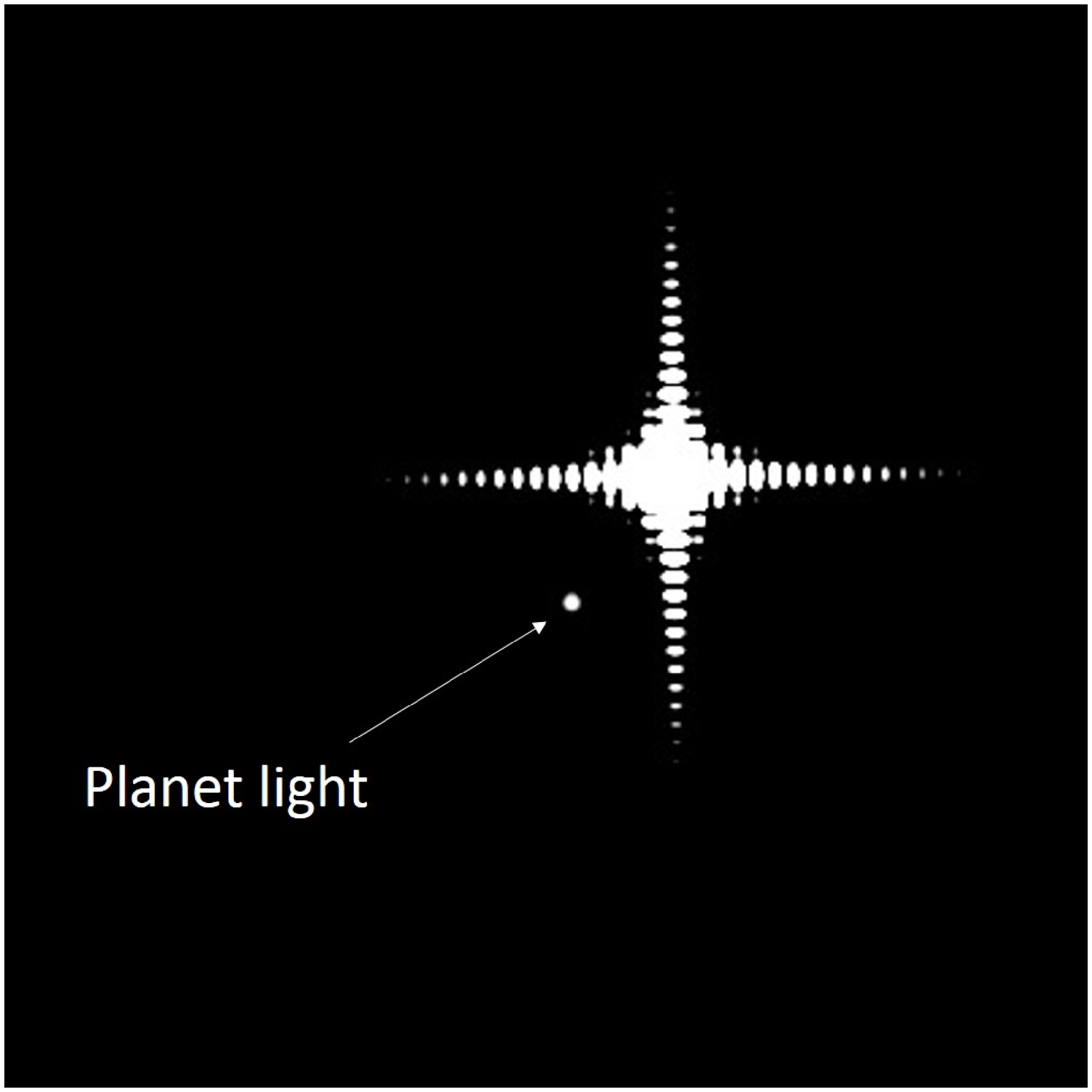}
   \caption{The intensity of starlight with the exposure of 0.5ms (left), the intensity of planet light with the exposure of 85ms (right).}
   \label{Fig5}
   \end{figure}
   \begin{figure}
   \centering
  \includegraphics[scale=0.72]{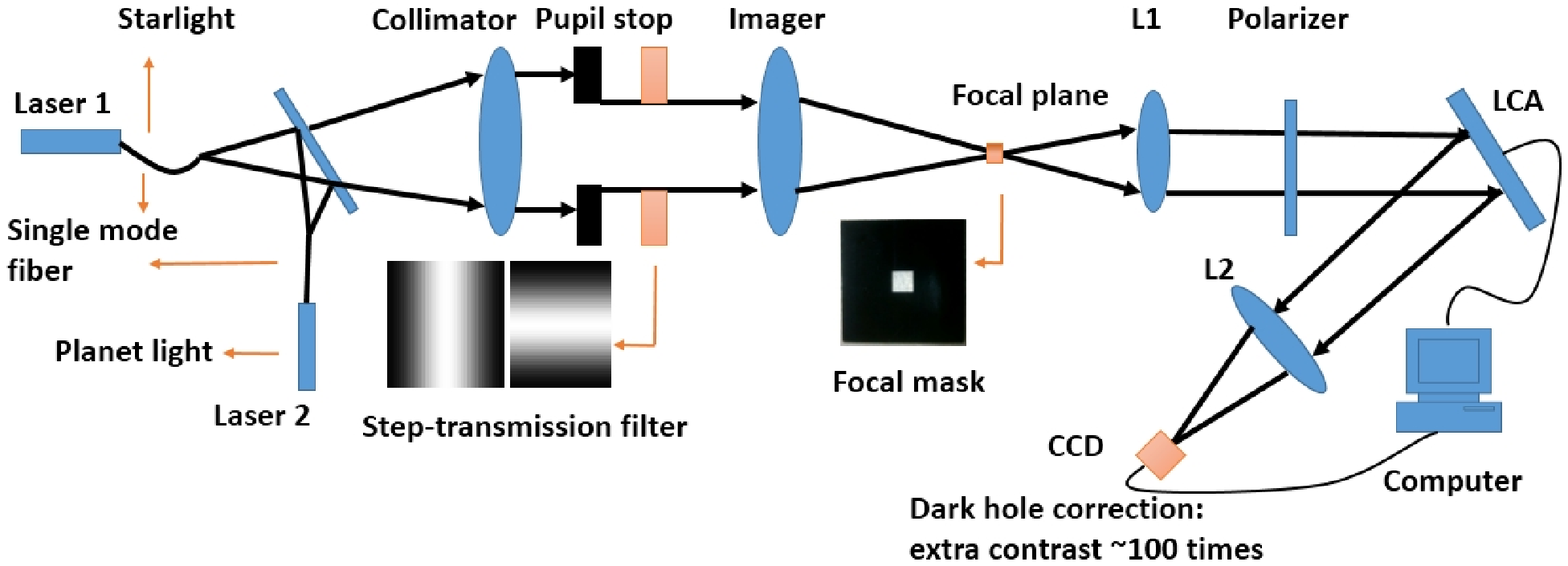}
   \caption{The whole optical system of dark-hole correction.}
   \label{Fig6}
   \end{figure}
\section{CONFIGURATIFig1ON OF DARK-HOLE CORRECTION OPTICS SYSTEM AND THE DARK HOLE CREATION}
\subsection{The intensity calibration of starlight and planet light}

Since a 16-bit CCD camera can only measure a contrast up to 1$/$65,536 (about $10^{-4}$) and the contrast we will measure is out of this range, we need use a neutral density (ND) filter to attenuate the star light. Two laser light sources (laser 1 and laser 2) are used to simulate the starlight and the planet light, respectively, with the optical configuration of intensity calibration shown in Figure 4. We need to know the contrast of the two beams. Therefore, before running the optimization code, we firstly calibrate the intensity of the starlight and planet light sources. The step transmission filters must be removed. A ND filter with 1$/$1000000 intensity attenuation is inserted in the star beam. In two individual exposures, we adjust the camera exposure time to make the light intensity in the planet and the star equal each other. Figure 5 shows the images in two different exposure times. According to the exposure times and ND attenuation, the contrast between the two beams should be 85$/$0.5$/$1000000$=$1.7$\times$$10^{8}$. Once the contrast of the two sources is known, it can be used to calculate a higher contrast of the coronagraph. When the intensity calibration is accomplished, we run the SPGD optimization code.
\subsection{Optical layout of the whole system}
Figure 6 shows the layout of the coronagraph optical system under test. The starlight and planet light are simulated by two point lights using two laser light sources ($\lambda=632.8nm$), who's contrast are calibrated in advanced. When we run the SPGD optimization code, the laser used to simulate the planet light will be blocked. The starlight is suppressed when the light passes through the pupil stop and two orthogonal step transmission filters

Due to the small number of DM actuators constraints, it is not possible to create a large dark hole. In addition, A DM with large actuators is expensive. Therefore, we use a LCA to replace a DM in our current test. In this test, only one quadrant is used for high contrast imaging. A focal mask composed of a square aperture, is located on the first focal plane (before lens L1), which blocks three quadrants of the focal plane image and only allows one quadrant image to go through the optical system for high-contrast imaging.

The LCA is located on a plane conjugate to the pupil. A polarizer should be placed before the LCA in order to control incident light polarization state, and the LCA is working as a phase-only modulator when the input light source is linearly polarized in its vertical axis. The phase of the incoming linear polarized light can be modulated by the LCA, as the light passes through it. Finally, the light is imaged by a lens (L2), and a fast CCD camera takes the PSF images.

The PSF images are taken by a 14-bit fast CCD camera and saved in computer in real-time. Then using PSF image the SPGD algorithm evaluates the energy in the target region of interest. According to the evaluation, the SPGD will control the voltage of each LCA effective pixel to provide an overall phase to minimize the energy in the target field of view. Through multiple iterations, the expected dark-hole can be generated. Finally the computer saves the LCA optimal voltages, which corresponds the associated phase to generate an extra contrast improvement. And the dark hole can be locked stably in the lab environment for at least 6 hours, which is long enough for Jupiter like exoplanets imaging.
\begin{figure}
   \centering
   \includegraphics[scale=0.24]{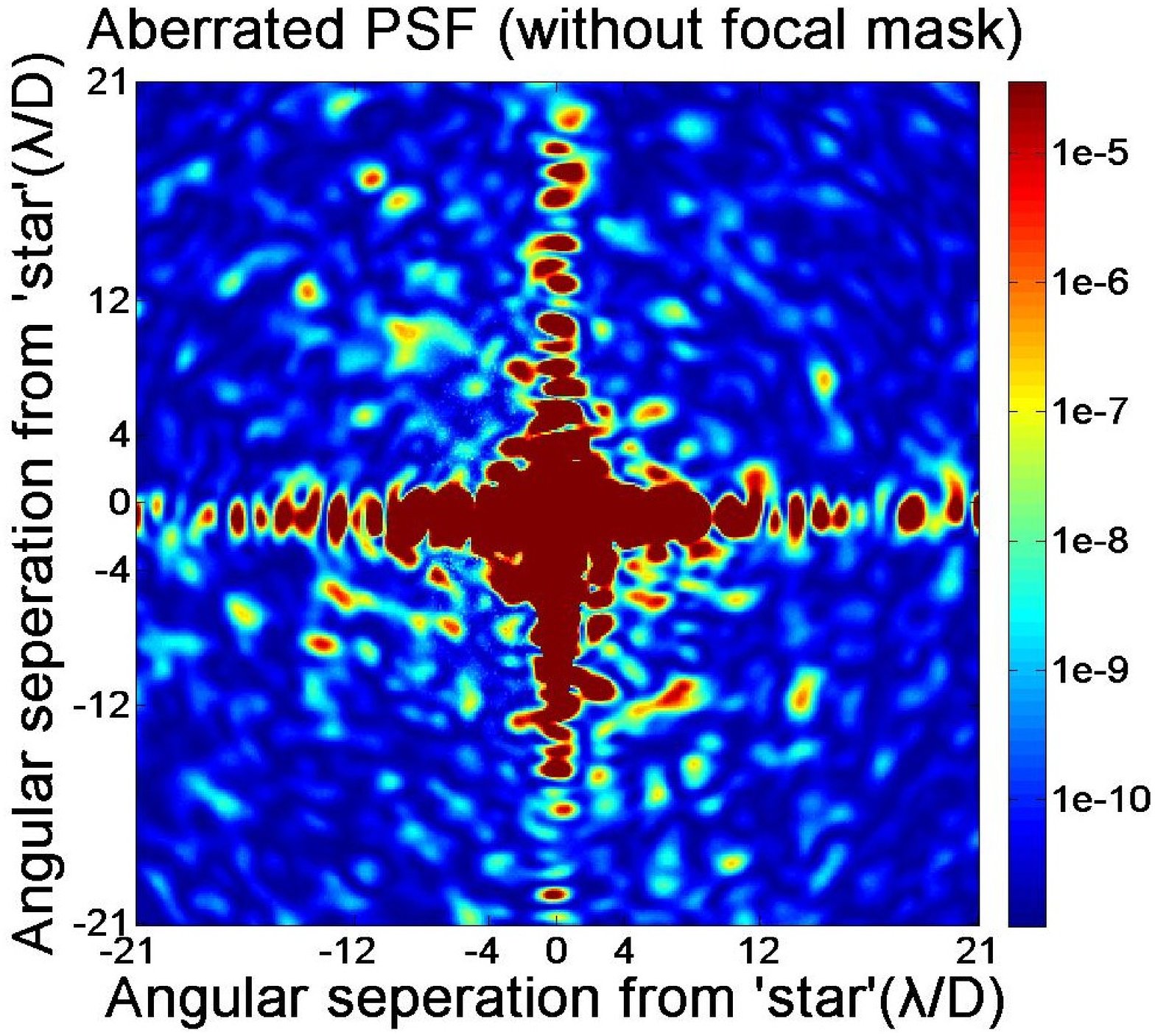}
   \includegraphics[scale=0.24]{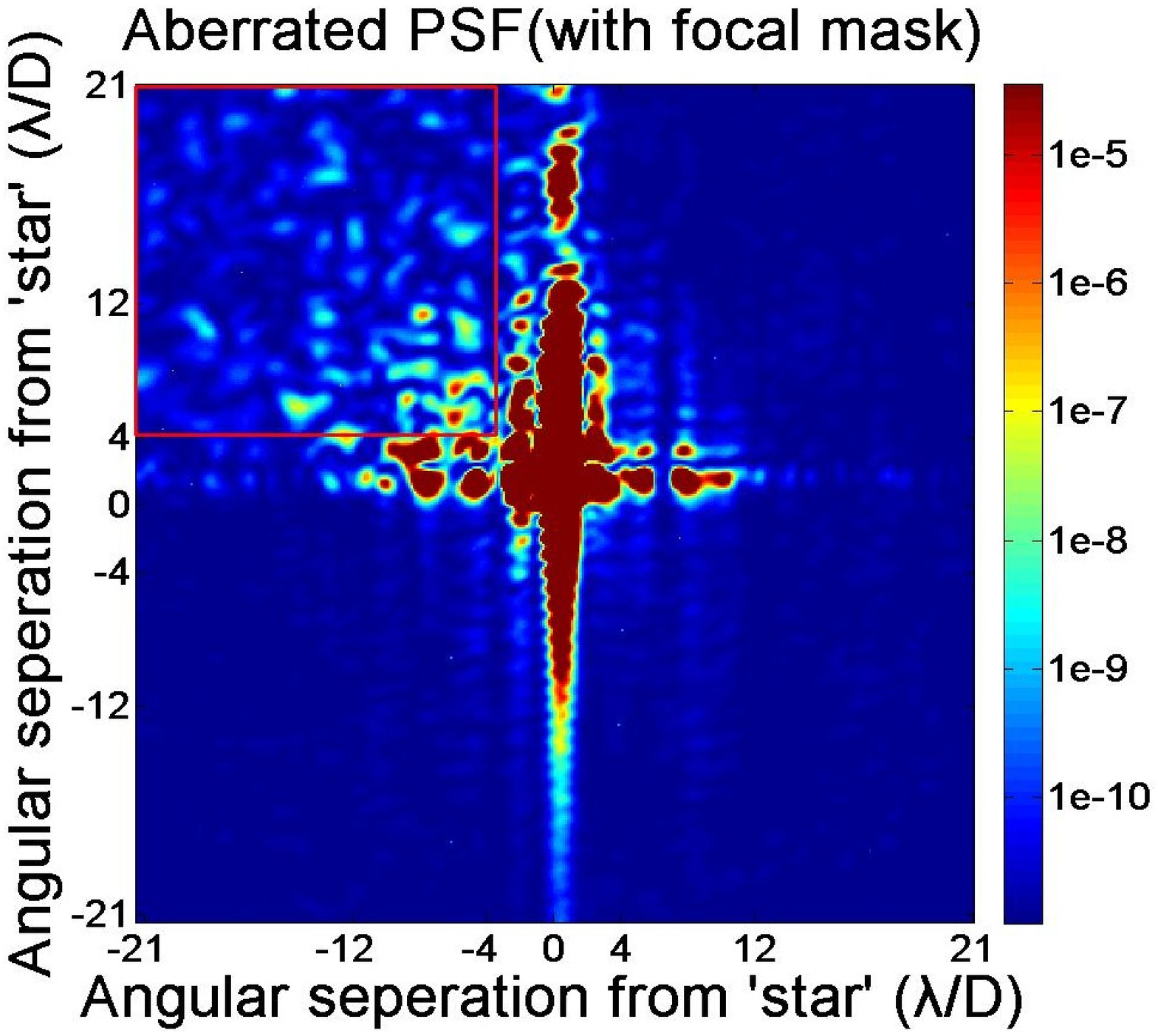}
   \includegraphics[scale=0.24]{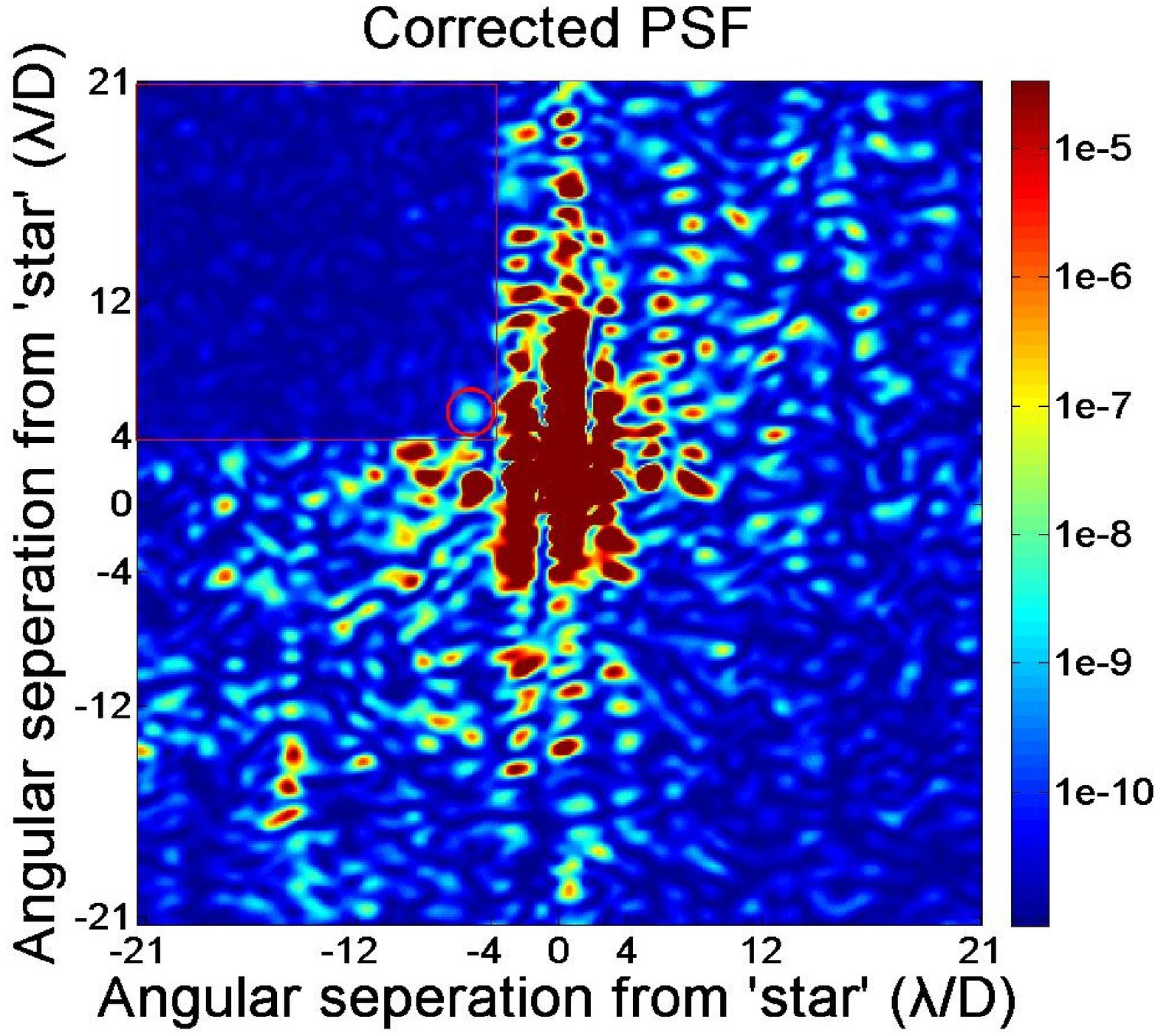}
   \caption{Aberrated and corrected PSFs in our testbed. The left panel shows the PSF without focal plane mask, while the central panel shows the PSF with focal mask. The focal plane image after the dark-hole correction is presented in the right panel. In the right panel, the planet image (indicated by a red circular rigion) stand out from the speckle noise in the red rectangle target region defided by the dark hole, with a region of $X = Y = 4\lambda/D-21\lambda/D$. This monochromatic experiment is done at the wavelength of 632.8nm.}
   \label{Fig7}
   \end{figure}
\subsection{Dark hole creation}
The dark hole obtained using our SPGD optimization algorithm is shown in Figure7. A dark-hole (between $4\lambda/D$ and $21\lambda/D$) is created in a quadrant on the focal plane that is larger than JPL's (between $4\lambda/D$ and $10\lambda/D$) and Princeton's ($X=7-10\lambda/D$ and $Y=-3-3\lambda/D$) (\citealt{Pueyo+etal+2009}). After 1000 iterations of the phase correction, the speckle noise in the original PSF has been efficiently removed. Once the dark-hole is created, we turn on the laser light and send the optimized LCA phase.Then the planet image stands out the speckle noise in the dark hole region, and can be seen clearly, as shown in the right panel of Figure 7. In fact the intensity of the planet is about 3.5 times brighter than residual speckle noise in the dark-hole. Therefore, the contrast in the dark hole increases from 1$\times$$10^{-7}$ to 1.68$\times$$10^{-9}$ at $4\lambda/D$, which demonstrates an extra gain in contrast of two orders of magnitudes through the contribution of the dark hole using the LCA.
\section{CONCLUSIONS AND FUTIRE WORK}
\label{sect:discussion}
In this recent test, we have presented the test results of our new 51-step transmission filters in our high-contrast coronagraph. According to the test, the contrast can reach $10^{-7}$ at $4\lambda/D$, without phase correction. Then we used the LCA to generate a dark-hole based on our SPGD algorithm with the new filters. In the dark-hole region, an extra contrast $\thicksim$100 times has been achieved. Therefore, our high-contrast coronagraph with the step transmission filters achieves a contrast of $10^{-9}$ in the dark-hole region. Our approach has following advantages: the wave-front control algorithm is simple and is done based on the focal plane PSF only; the dark-hole region created is larger, which does not need high density actuator DMs.

This paper demonstrates that our coronagraph is a promising technique for space-based imaging earth-like planets. For future works, we will use a DM ($32 \times 32$ or $64 \times 64$ actuators) whose precision ($\lambda/10000$) for phase correction is better than that of the LCA ($\lambda/100$), and the experiment will be carried out in a space-like environment inside a vibration-isolated and vacuum chamber. Based on the two points, the contrast should be able to reach or better than $1\times10^{-10}$. Further progresses will be discussed in our future publications.
\normalem
\begin{acknowledgements}
This work was supported by the National Natural Science Foundation of China (Grant Nos. 11220101001, 11373005 and 11328302) and the "Strategic Priority Research Program" of the Chinese Academy of Sciences (Grant No. XDA04070600 and XDA04075200). This work was made possible through the support of a grant from the John Templeton Foundation and National Astronomical Observatories of Chinese Academy of Sciences. Part of the work described in this paper was carried out at California State University Northridge, with support from the National Science Foundation under Grant ATM-0841440.
\end{acknowledgements}

\bibliographystyle{raa}
\bibliography{bibtex}

\begin{thebibliography}{11}
\providecommand{\natexlab}[1]{#1}
\providecommand{\selectlanguage}[1]{\relax}

\bibitem[{Borucki et~al.(2013)Borucki, Agol, Fressin
  et~al.}]{Borucki+etal+2013}
Borucki, W.~J., Agol, E., Fressin, F., et~al. 2013, Science, 340, 587

\bibitem[{Dong et~al.(2011)Dong, Ren, \& Zhang}]{Dong+etal+2011}
Dong, B., Ren, D.-Q., \& Zhang, X. 2011, RAA(Research in Astronomy and
  Astrophysics), 11, 997

\bibitem[{Dou et~al.(2009)Dou, Ren, Zhu, \& Zhang}]{Dou+etal+2009}
Dou, J., Ren, D., Zhu, Y., \& Zhang, X. 2009, in SPIE Optical Engineering+
  Applications, 744019--744019 (International Society for Optics and Photonics)

\bibitem[{Dou et~al.(2010)Dou, Ren, \& Zhu}]{Dou+etal+2010}
Dou, J.-P., Ren, D.-Q., \& Zhu, Y.-T. 2010, RAA(Research in Astronomy and
  Astrophysics), 10, 189

\bibitem[{Kay et~al.(2009)Kay, Pueyo, \& Kasdin}]{Kay+etal+2009}
Kay, J.~D., Pueyo, L.~A., \& Kasdin, N.~J. 2009, in SPIE MOEMS-MEMS: Micro-and
  Nanofabrication, 72090G--72090G (International Society for Optics and
  Photonics)

\bibitem[{Pepe et~al.(2013)Pepe, Cameron, Latham et~al.}]{Pepe+etal+2013}
Pepe, F., Cameron, A.~C., Latham, D.~W., et~al. 2013, Nature, 503, 377

\bibitem[{Pueyo et~al.(2009)Pueyo, Kay, Kasdin et~al.}]{Pueyo+etal+2009}
Pueyo, L., Kay, J., Kasdin, N.~J., et~al. 2009, Appl. Opt, 48, 6296

\bibitem[{Ren et~al.(2012)Ren, Dong, Zhu, \& Christian}]{Ren+etal+2012}
Ren, D., Dong, B., Zhu, Y., \& Christian, D.~J. 2012, PASP, 124, 247

\bibitem[{Ren et~al.(2010)Ren, Dou, \& Zhu}]{Ren+etal+2010}
Ren, D., Dou, J., \& Zhu, Y. 2010, PASP, 122, 590

\bibitem[{Ren \& Zhu(2007)}]{Ren+Zhu+2007}
Ren, D., \& Zhu, Y. 2007, PASP, 119, 1063

\bibitem[{Trauger \& Traub(2007)}]{Trauger+Traub+2007}
Trauger, J.~T., \& Traub, W.~A. 2007, Nature, 446, 771

\end{thebibliography}


\begin{thebibliography}{0}
\providecommand{\natexlab}[1]{#1}
\providecommand{\selectlanguage}[1]{\relax}

\end{thebibliography}

\end{document}